\documentclass[aps,prd,amssymb,showpacs,floatfix,twocolumn,superscriptaddress,nofootinbib]{revtex4-1}
\usepackage{amsmath,amsfonts,graphics,epsfig}

\begin{document}

\title{Magnetic moment of the $\rho$ meson in instant-form
relativistic quantum mechanics}
\author{A.F.~Krutov}
\email{krutov@ssau.ru}, \affiliation{Samara University, 443086 Samara,
Russia} \author{R.G.~Polezhaev} \email{polezaev@list.ru},
\affiliation{Samara University, 443086 Samara, Russia}
\author{V.E.~Troitsky} \email{troitsky@theory.sinp.msu.ru}
\affiliation{D.V.~Skobeltsyn Institute of Nuclear Physics,\\
M. V. Lomonosov Moscow State University, Moscow 119991, Russia}
\date{\today}
\begin{abstract}

We derive an explicit analytical expression for the
magnetic dipole moment of the $\rho$ meson, $\mu_{\rho}$,
in a relativistic constituent-quark model.
We adopt our relativistic approach to composite systems, modified
instant-form (mIF)
Relativistic  Quantum Mechanics (RQM),
that we used
particularly to construct
a unified $\pi $\&$\rho $ model
 (Phys. Rev. D \textbf{93}, 036007 (2016))
describing electroweak properties of
light mesons.
 This model provides a parameter-free
calculation to give
$\mu_{\rho} = 2.16 \pm 0.03$~\hspace{1mm}$[e/2M_{\rho}]$
which is in accordance with the
conventional experimental data. The magnetic, quadrupole and charge form
factors also are derived and presented. We consider the small uncertainty
of our value of magnetic moment as one of undoubted advantages of the
method. A comparison is made with recent lattice QCD results and previous
calculations using a variety of methods.

\end{abstract}

\maketitle

\section{Introduction}
\label{sec: introduc}
For our understanding of the structure of strong interaction
the electromagnetic form factors
are of fundamental importance and give complementary information.
Hadron form factors provide an important tool for understanding
the structure of bound states in Quantum Chromodynamics (QCD). The study of
electromagnetic properties of the $\pi$ and $\rho$ mesons consisting of
two light ($u, d$) quarks is of particular interest because in numerous
respects they are the simplest bound states.

It is worth noting that the experimental situation is
quite different in the cases of $\pi$ and
$\rho$ mesons.
The pion properties are well known from experiments.
On the contrary,
the experimental data on the $\rho $ meson are scarce.
Its
lifetime is very short, $\sim 4.5\times10^{-24}$~s, so direct
measurements of its electroweak properties (e.g., electromagnetic
form factors and static moments) are nearly impossible.

Nevetheless, these form factors are important for hadron physics: for
example,  they contribute to meson exchange currents. They are also closely
related to axial-vector diquark models of nucleon form factors. Besides,
 the description of the $\rho$ meson is of interest on its own account.
Under such conditions, the role of theoretical investigations does
increase. Indeed, recently the  interest in $\rho$ meson
form factors has been renewed. During last years a number of theoretical
approaches to the structure of the $\rho $ meson appeared although their
results cannot be directly compared to measurements. Below we review these
approaches while discussing the results of the calculations presented in
the Table 1.

Theoretical approaches to the description of
bound states are split into two directions. From the high-energy side,
QCD, which is widely believed to be a fundamental
theory of the strong forces, becomes strongly coupled at the relevant
energy scales, so trustable perturbative calculations help a little in the
quantitative description of precise low-energy  data, which there is no
lack of. From the low-energy side, a number of successful models to
describe the data have been developed. To be quantitative, they
necessarily require some phenomenological input. None of these models can
be consistently and quantitatively derived from the QCD Lagrangian.

The approach that we use in the present paper is a particular
relativistic formulation
of constituent-quark model that is based on the classical paper
by P.~Dirac \cite{Dir49} (so-called Relativistic Hamiltonian
Dynamics or
Relativistic  Quantum Mechanics (RQM)). RQM can be formulated in different
ways or in different forms of dynamics. The main forms are instant form
(IF), point form (PF) and light-front (LF) dynamics. The description of
different forms of RQM dynamics can be found in
the reviews \cite{LeS78,KeP91,Coe92,KrT09}. Today the approach is
largely used for nonperturbative description of particle structure. It
gives an opportunity to construct a nonperturbative QCD model on the basis
of constituent-quark models. It is just such a kind of model that we have
constructed \cite{KrP16} --
the unified $\pi $\&$\rho $ model
with no free parameters remaining.

Now our goal is to obtain the value of the $\rho$ meson magnetic moment
$\mu_{\rho}$  using explicit analytic formula which we derive here within
the unified $\pi $\&$\rho $ model.
It is extremely important to emphasize that our $\mu_{\rho}$
is obtained without any fitting parameters: in
the unified $\pi $\&$\rho $ model \cite{KrP16}
there remains no possibility to fit, all parameters are already fixed.
Our result for the magnetic moment,
$\mu_{\rho} = 2.16 \pm
0.03$~\hspace{1mm}$[e/2M_{\rho}]$, compares well with the
conventional experimental result, but with significantly smaller
uncertainty.

The model constructed in \cite{KrP16} is based on our version of
RQM, -- the modified instant form (mIF) of RQM \cite{KrT02},
\cite{KrT03}, and on the actual calculation of the $\pi$-meson structure
\cite{KrT01} in 1998. This model has predicted with surprising accuracy the
values of $F_{\pi}(Q^{2})$, which were measured later in JLab experiments
\cite{Vol01,Hor06,Tad07,Blo08,Hub08} (see discussion in Ref.
\cite{KrT09prc}).
Another advantage of the approach is matching with the QCD predictions in
the ultraviolet limit: when consistent-quark masses are switched off, as
expected at high energies, the model reproduces correctly not only the
functional form of the QCD asymptotics, but also the numerical
coefficient; see Refs. \cite{KrT98,TrT13,TrT15} for details. Let us note
that mIF was successfully used also  for other composite systems, namely,
the deuteron \cite{KrT03eur,KrT07,KrT08} and the $K$-meson \cite{KrT17}.

The rest of paper is organized as follows. In
Sec. \ref{sec:mod} we recall briefly the main points of our model.
The $\rho$ meson electromagnetic current is
written in terms of the Sachs form factors as well as in terms of
form factors appropriate to the general method of relativistic invariant
parametrization of matrix elements of local operators that we use.
The brief description of the modified impulse approximation (MIA) is given.
In Sec. \ref{sec:res} the $\rho$-meson magnetic
moment is obtained. It is shown that the electromagmetic static moments
of the $\rho$ meson
are
to be considered as
regular generalized functions (distributions)
determined on the space of test functions -- quark-quark wave
functions.
To calculate the static moments one has to take the weak limits
at $Q^2\to 0$
of corresponding functionals. The explicit analytic
expression for the $\rho$ meson magnetic moment is obtained.
We discuss the
values of the parameters of the model which were defined previously in
\cite{KrP16}, and that we use here. The value of the calculated
$\rho$ meson magnetic moment is given. This Section contains
a discussion and the comparison of obtained results with the results of
other authors. We briefly conclude in Sec.\ref{sec:concl} and present
some details of the calculation in the Appendix.

\section{Electromagnetic form factors of the $\rho$ meson in modified
instant form
RQM}
\label{sec:mod}

In this paper we use  the relativistic constituent
model that describes the hadron properties at the quark level
in terms of degrees of freedom of constituent quarks. The
constituent quarks are considered as extended objects, the
internal characteristics of which (mean square radius, anomalous
magnetic moments, form factors) are parameters of the
model. As a relativistic variant of the constituent model we
choose the method of RQM.

The RQM method, as a relativistic theory of composite
systems, is based on the direct realization of the Poincar\'e
algebra on the set of dynamical observables on the Hilbert
space. The RQM theory of particles lies between local field
theoretic models and nonrelativistic quantum mechanical
models.

Contrary to the field theory, RQM deals with a finite number
of degrees of freedom from the very beginning. This is certainly
a kind of a model approach. The preservation of the
Poincar\'e algebra ensures the relativistic invariance. So, the
covariance of the description in the frame of RQM is due to
the existence of the unique unitary representation of the inhomogeneous
group $SL(2,C)$ on the Hilbert space of composite
system states with a finite number of degrees of freedom.

The mathematics of RQM is similar to that of nonrelativistic
quantum mechanics and permits one to assimilate the
sophisticated methods of phenomenological potentials. It
can be generalized to describe three or more particles. RQM is
based on the simultaneous action of two fundamental principles,
relativistic invariance and the Hamiltonian principle,
and presents the most adequate tool to treat the systems with a
finite number of degrees of freedom.
The use of RQM enables one
to separate the main degrees of freedom and thus to construct
a convenient relativistic invariant approach to
the electroweak structure of two-particle composite systems.

We use one of the forms of RQM, namely a version of the
IF. The dynamics of a composite system, that is the interaction of the
constituents is described in IF in the frame of the general RQM axiomatics.
This means, particularly, that the inclusion of the interaction in the
algebra of  the Poincar\'e group is realized by means of the
Bakamjian-Thomas procedure
\cite{BaT53} which permits to preserve the commutation relations between
the generators of the group (the observables
(see the reviews \cite{KeP91,KrT09}).
Following this procedure one includes the constituent-interaction
operator by adding it to the operator of the mass of the free constituent
system:
\begin{equation}
\hat M_0
\to \hat M_I = \hat M_0 + \hat V \;.
\label{M0toMI}
\end{equation}
Here $\hat M_0$ is the operator of the invariant mass of the system
without interaction and
$\hat M_I$ is the operator of the mass of interacting system.
In IF dynamics, the interaction operator obeys the following conditions:
\begin{equation}
\hat M_I = \hat M_I^+\;,\quad \hat M_I\;>\;0\;,
\label{MI+}
\end{equation}
\begin{equation}
\left [\hat {\vec P},\,\hat V\right ] = \left
[\hat {\vec J},\,\hat V\right ] = \left [
\vec\bigtriangledown_P,\,\hat V\right ] = 0\;.
\label{[PU]=0}
\end{equation}
The conditions (\ref{MI+}) present spectral conditions for the mass
operator. The equations
(\ref{[PU]=0}) ensure that the algebraic relations of the Poincar\'e
algebra are  fulfilled. The constraints  (\ref{[PU]=0}) are not too
restrictive. For example, they are satisfied by all nonrelativistic
interaction potentials. The equations  (\ref{[PU]=0}) mean, in particular,
that the interaction potential does not depend on the total moment of the
system.
In RQM the wave function of the system of interacting
particles is defined as the eigenfunction of a complete set of commuting
operators. In IF this set is:
\begin{equation}
 {\hat M}_I^2\;(\hbox{or}\;\hat M_I)\;,\quad
{\hat J}^2\;,\quad \hat J_3\;,\quad \hat {\vec P}\;.
\label{complete}
\end{equation}
${\hat J}^2$
is the operator of the square of the total angular momentum.
In IF the operators
${\hat J}^2\;,\;
\hat J_3\;,\; \hat {\vec P}$
coincide with
those for the free system. So, in (\ref{complete})
 only the operator $\hat M_I^2\;(\hat M_I)$
depends on the interaction.

To diagonalize the operators ${\hat J}^2\;,\;\hat J_3\;,\; \hat {\vec P}$
one has
first to construct the adequate basis in the state space of
composite system. In the case of two-particle system
the Hilbert space
in RQM is the direct product of two one-particle Hilbert
spaces:
${\cal H}_{q\bar q}\equiv {\cal
H}_q\otimes {\cal H}_{\bar q}$.
As a basis in
${\cal
H}_{q\bar q}$
 one can choose the following set of two-particle state vectors
where the motion of the
center of mass is separated:
\begin{equation}
|\,\vec P,\;\sqrt {s},\;J,\;l,\;S,\;m_J\,\rangle\;.
\label{PkJlSm}
\end{equation}
Here
$P_\mu = (p_1 +p_2)_\mu$, $P^2_\mu = s$, $\sqrt {s}$
is the invariant
mass of the two-particle system, $l$ is the orbital angular
momentum in the center-of-mass frame (C.M.S.),
$\vec S\,^2=(\vec S_1 +\vec S_2)^2 = S(S+1)\;,\;S$
is the total spin in
C.M.S.,$J$ is the total angular momentum with the projection
$m_J$, and
$p_1\;,\;p_2$ are the constituent moments.
The two-particle basis with separated motion of the center of mass
(\ref{PkJlSm}) is connected with the basis of individual spins and
momenta of two particles
through the appropriate
Clebsh-Gordan  decomposition
 for the Poincar\'e group
(see, e.g., \cite{KrT09}).

So, to obtain the system wave function in the basis
(\ref{PkJlSm}) one needs to diagonalize the operator
$\hat M_I^2$ (or $\hat M_I$). The eigenvalue problem for the
operator $\hat M_I^2$  can be written in
the form of nonrelativistic
Schr\"odinger equation, the corresponding interaction operator having the
meaning of a phenomenological nonrelativistic potential.
The two-particle wave function of relative motion for
fixed total angular momentum is:
\begin{equation}
\varphi^{J}_{lS}(k(s)) =\sqrt[4]{s}\,u_{lS}(k)\,k\;,\quad
\sum_{lS}\int\,u_{lS}^2(k)\,k^2\,dk = 1\;,
\label{phi(s)}
\end{equation}
where $u_{lS}(k)$  is a model wave function which is a solution of the
eigenvalue problem for the operator $\hat M_I^2$ or $\hat M_I$ in the
representation given by the basis
(\ref{PkJlSm}), $k = \sqrt{s-4\,M^2}/2$. The normalization
factors that stay in (\ref{phi(s)}) with
$u_{lS}(k)$ correspond to the transition to the relativistic
density of states:
\begin{equation}
k^2\,dk\quad \to\quad \frac{k^2\,dk}{2\sqrt{k^2 + M^2}}\;.
\label{rel den}
\end{equation}

It is worth to notice that wave functions in RQM defined as the
eigenfunctions of the operators set  (\ref{complete}) in general are not
the same as relativistic covariant wave functions defined as solutions of
wave equations or as the matrix elements of a local Heisenberg field.

Our approach has a number of features that distinguish it
from other forms of dynamics and other approaches in the
frames of IF \cite{KrT09, KrT02, KrT03}. In particular, this approach (see,
for example, \cite{KrT09, KrT03}) differs essentially from that of
traditional RQM in what concerns the method of construction of transition
current operators. The main point of our approach to the construction  of
the electroweak current operator is the so-called method of the canonical
 parametrization of the local operator matrix elements. The foundation of
the method was given in \cite{ChS63} and it was generalized to the case of
composite systems in \cite{TrS69, KoT72, KrT05}. This parametrization is a
realization of the Wigner-Eckart theorem for the Poincar\'e group
\cite{KrT05} and so it enables one, for given matrix element of arbitrary
tensor dimension, to separate the reduced matrix elements (form factors)
that are invariant under the action of the Poincar\'e group. Matrix
element of an operator is presented by a sum of terms which are the
products of a covariant and an invariant parts. The covariant part of the
matrix element describes its transformation (geometrical) properties while
all the dynamical information about the transition
 is contained in the invariant part
-- in the reduced matrix elements, or form factors.
In our approach, the reduced matrix elements are generalized functions.
Strictly speaking, they are generalized functions in all cases of
composite models, in particalar, if one uses impulse approximation.

Our approach has the following characteristic features:

(a) The electroweak current matrix element satisfies automatically
the relativistic covariance conditions and in the case of the
electromagnetic current also the conservation law.

(b) We propose a modified impulse approximation (MIA), which is formulated
in terms of reduced matrix elements
and not in terms of operators as it takes place in standard
impulse approximation. It is constructed in a relativistic
invariant way. This means that our MIA does not depend on the choice of
the coordinate frame, and this contrasts principally with the
``frame-dependent'' impulse approximation usually used in the instant form
(IF) of dynamics.

(c) For composite systems (including the spin-1 case) the
approach guarantees the uniqueness of the solution for form
factors and does not use such concepts as ``good'' and ``bad''
current components.

The analytic properties of the pion form factor in the complex plane of the
transfer momentum square in our model correspond to properties that follow
from general principles of  quantum field theory
\cite{KrN13}.  The
model was also applied to the calculation of electroweak parameters of the
$\rho$ meson \cite{KrP16}, for which particularly interesting relations
have been obtained. It is just those results which give the parameter-free
method that permits to obtain the magnetic moment of the $\rho$ meson
without any fitting.

The static electromagnetic moments of a particle are the limit values of
Sachs form factors at $Q^2\;\to\;0$.
For spin-1 particles, including
$\rho$ meson, the electromagnetic current matrix element can be written in
terms of Sachs form factors in the Breit frame as follows
(see, e.g., \cite{ArC80,BrH92}):
$$
\langle\vec p_\rho\,,m_J|j_\mu(0)|\vec p_\rho\,'\,,m'_J\rangle =
G^\mu(Q^2)\;,
$$
$$
G^0(Q^2) =
2p_{\rho0}\left\{(\vec\xi\,'\vec\xi\,^*)\,G_C(Q^2)\right. +
$$
$$
\left. \left[(\vec\xi\,^*\vec Q)(\vec\xi\,'\vec Q) -
\frac{1}{3}Q^2
(\vec\xi\,'\vec\xi\,^*)\right]\,\frac{G_Q(Q^2)}{2M_\rho^2}\right\}\;,
$$
\begin{equation}
\vec G(Q^2) = \frac{p_{\rho
0}}{M_\rho}\left[\vec\xi'\,(\vec\xi\,^*\vec Q) -
\vec\xi\,^*(\vec\xi\,'\vec Q)\right]\,G_M(Q^2)\;. \label{Grho}
\end{equation}
Here $G_C\;,\;G_Q\;,\;G_M$  are the charge, quadrupole and
magnetic form factors, respectively,
$$
q^\mu = (p_\rho - p'_\rho)^\mu = (0\;,\;\vec Q)\;,
$$
$$
p_\rho^\mu = (p_{\rho0}\;,\;\frac{1}{2}\vec Q)\;,\quad
p'_\rho\,^\mu = (p_{\rho0}\;,\;-\frac{1}{2}\vec Q)\;,
$$
$$
p_{\rho0} = \sqrt{M_\rho^2 + \frac{1}{4}Q^2}\;,\quad \vec Q =
(0\;,\;0\;,\;Q)\;.
$$
$$
\xi^\mu(\pm 1) = \frac{1}{\sqrt{2}}(0\;,\;\mp
1\;,\;-\,i\;,\;0)\;,\; \xi^\mu(0) = (0\;,\;0\;,\;0\;,\;1)\;.
$$
The arguments of the polarization vector
$\xi$ are the projections of the total  angular momentum.
The static limits of form factors in (\ref{Grho}) give the magnetic
$\mu_\rho$ (in units $e/2M_{\rho}$) and the quadrupole $Q_\rho$ moments
\cite{ArC80}:
\begin{equation}
G_M(0) = \mu_\rho\;,\quad G_Q(0) =
M_\rho^2\,Q_\rho\;. \label{stat}
\end{equation}

One can use the general procedure of relativistic covariant construction
of local operators matrix elements to obtain, in our version of IF RQM,
the current matrix element (\ref{Grho}) in an arbitrary coordinate system
\cite{KrT03}:
$$
\langle\vec p_c\,,m_{J\rho}|j_\mu(0)|\vec
p_\rho\,'\,,m'_{J\rho}\rangle =
$$
$$
\langle\,m_{J\rho}|\,D^{1}(p_\rho\,,p'_\rho)\, \sum_{i=1,3}\,
\tilde{\cal F}\,^i_\rho(t)\,\tilde A^i_\mu\,|m'_{J\rho}\rangle\;,
$$
$$
\tilde{\cal F}\,^1_c(t) = \tilde f^\rho_{10} + \tilde
f^\rho_{12}\left\{[i{p_\rho}_\nu\,\Gamma^\nu(p'_\rho)]^2 \right.
\left. \right. -
$$
\begin{equation}
\left.
\frac{1}{3}\,\hbox{Sp}[i{p_\rho}_\nu\,\Gamma^\nu(p'_\rho)]^2\right\}
\frac{2}{\hbox{Sp}[{p_\rho}_\nu\,\Gamma^\nu(p'_\rho)]^2}\;,
\label{fin}
\end{equation}
$$
\tilde{\cal F}\,^3_\rho(t) = \tilde f^\rho_{30}\;,
$$
$$
\tilde A^1_\mu = (p_\rho + p'_\rho)_\mu\;,\quad \tilde A^3_\mu =
\frac{i}{M_\rho} \varepsilon_{\mu\nu\lambda\sigma}
\,p_\rho^\nu\,p'_\rho\,^\lambda\,\Gamma^\sigma(p'_\rho)\;.
$$
Here $p'_\rho, p_\rho$ are 4-momenta of  the $\rho$ meson in initial
and final states, respectively, $m'_{J\rho}\,,m_{J\rho}$ are
projections of the total angular momenta,
$D^{1}(p_\rho\,,\,p'_\rho)$ is the matrix of Wigner rotation, $M_\rho$
is the $\rho$ meson mass, $\tilde f^\rho_{10}\,,\,\tilde
f^\rho_{12}\,,$ $\tilde f^\rho_{30}$ are the charge, quadrupole and
magnetic form factors of the $\rho$ meson, respectively.

The spin 4-vector
$\Gamma^\nu(p_\rho)$ is (see, e.g., \cite{KrT09}):
$$
\Gamma_0(p_\rho) = (\vec p_\rho\vec j)\;,\quad \vec \Gamma(p_\rho)
= M_\rho\,\vec j + \frac {\vec p_\rho(\vec p_\rho\vec
j)}{{p_\rho}_0 + M_\rho}\;,
$$
\begin{equation}
\quad \Gamma^2 = -M_\rho^2\,j(j+1)\;. \label{ Gamma mu}
\end{equation}

One can easily obtain the relation between the form factors in the form
(\ref{fin})
and Sachs form factors
(\ref{Grho}):
$$
G_C(Q^2) = \tilde f^\rho_{10}(Q^2)\;,\quad G_Q(Q^2) =
\frac{2\,M_\rho^2}{Q^2}\,\tilde f^\rho_{12}(Q^2)\;,
$$
\begin{equation}
\quad G_M(Q^2) = -\,M_\rho\,\tilde f^\rho_{30}(Q^2)\;.
\label{G=f}
\end{equation}

In the paper \cite{KrT03} the form factors
(\ref{fin}) were presented in the form of double integrals:
\begin{equation}
\tilde f^\rho_{in}(Q^2) =  \int\,d\sqrt{s}\,d\sqrt{s'}\,
\varphi(s)\,\tilde G_{in}(s,Q^2,s')\varphi(s')\;,
\label{intfin}
\end{equation}
where $\varphi(s)$ is the quarks wave function in the $\rho$ meson in the
sense of RQM, $\tilde G_{in}(s,Q^2,s')$ are the reduced matrix elements on
the Poincar\'e group. They are Lorentz-invariant regular generalized
functions.

It is worth to note that while obtaining
(\ref{intfin}) no assumption or approximation concerning the form of
the electromagnetic current operator was made. We have not
assumed, in particular, that the current operator is a sum of the
one-particle current operators of the costituents, that means that we have
not used the so-called impulse approximation which is known to break the
Lorentz-covariance and the conservation law for the composite-system
electromagnetic current in IF RQM. So, for
(\ref{intfin}) the Lorentz-covariance and current conservation law
are valid.

In general, the explicit form of the functions $\tilde G_{in}(s,Q^2,s')$,
is unknown. To calculate these functions
we propose a modified impulse approximation (MIA).
 In contrast to the generally accepted impulse
approximation
MIA is formulated
in terms of reduced matrix elements on the Poincar\'e group (form factors)
extracted from the current matrix element and not in terms of
current operators themselves.
For the case (\ref{intfin}) MIA means that the reduced matrix elements
$\tilde G_{in}(s,Q^2,s')$ are changed for the  free two-particle
form factors with no interaction between components.
Such free form factors enter the electromagnetic current matrix
element for a system of two free particles with $\rho$-meson quantum
numbers. The corresponding formulae for these form factors are given
in the Appendix. Note that the current
matrix element
as a whole still contains some contributions of two-particle currents in
a way which
such that ensures its correct transformation properties.

Using the relations (\ref{G=f})
we derive the integral representations for the $\rho$-meson
Sachs form factors in MIA:
$$
G_C(Q^2) =\int\,d\sqrt{s}\,d\sqrt{s'}\,
\varphi(s)\,g_{0C}(s\,,Q^2\,,s')\, \varphi(s')\;,
$$
\begin{equation}
G_Q(Q^2) =\frac{2M_\rho^2}{Q^2}\int d\sqrt{s} d\sqrt{s'}
\varphi(s) g_{0Q}(s\,,Q^2\,,s')\varphi(s')\;,
\label{GqGRIP}
\end{equation}
$$
G_M(Q^2) =-\,M_\rho\,\int\,d\sqrt{s}\,d\sqrt{s'}\,
\varphi(s)\,g_{0M}(s\,,Q^2\,,s')\, \varphi(s')\;,
$$
where
$g_{0C},\;g_{0Q},\;g_{0M}$ are the free two-particle charge, quadrupole and magnetic form factors
for the system of two free fermions with quantum numbers of
$\rho$ meson respectively (the explicit expression for these form factors are given in Appendix),
$\varphi(s)$ is the two-quark $\rho$-meson wave function in the sense of
RQM,
\begin{equation}
\varphi(s) = \sqrt[4]{s}\,\psi(k)\,k\;,\quad k =
\frac{1}{2}\sqrt{s - 4\,M^2}\;\label{wfRQM}
\end{equation}
with the normalization condition:
$$
\int\,k^2\,\psi^2(k)\,dk\ = 1.
$$

As was shown in \cite{KrT02} the free two-particle electromagnetic form
factor is a regular generalized function (distribution)
defined on the space of test functions in {\bf R}$^2$,
$Q^2$ is the parameter of the generalized function.
This generalized function giving the $\rho$ meson form factor is given by the
functional $\langle g_{0i}(s,Q^2,s')\,,\phi(s,s')\rangle ,\,i=C,Q,M$ that is defined
as an integral of the product of the function $g_{0i}(s\,,Q^2\,,s')$ and
the test function $\phi(s,s')$. If we take the product of quark wave
functions in the initial and the final states for the test function, then
the considered functional is the corresponding $\rho$-meson form factor.

\section{The $\rho$-meson magnetic moment in $\pi $\&$\rho $ model.
Results and discussion}
\label{sec:res}

To calculate the $\rho$-meson magnetic moment in our approach we
have to take the static limit of the form factor (\ref{GqGRIP})
and to choose the values of the parameters entering our equations.
The limit is to be taken in the
weak sense and there is no need in fitting the parameters: all of them
were fixed in our $\pi $\&$\rho $ model in our previous work \cite{KrP16}.

It is easy to see that
the function (\ref{GqGRIP}) has no strong limit at $Q^2\;\to\;0$.  As the
function is a generalized function one has to take the limit in the weak
sense. So, the (weak) static limit of (\ref{GqGRIP}) gives the following
$\rho$-meson magnetic moment:

$$
\mu_\rho =
\frac{M_\rho}{2M}\,\int_{2M}^\infty\,d\sqrt{s}\,\frac{\varphi^2(s)}
{\sqrt{s - 4\,M^2}}\,\left\{1 - L(s) \right.
$$
\begin{equation}
\left. + (\kappa_u + \kappa_{\bar d}) \left[1 -
\frac{1}{2}\,L(s)\right]\right\}\;, \label{mu}
\end{equation}
$$
L(s) = \frac{2\,M^2}{\sqrt{s - 4\,M^2}\,(\sqrt{s} + 2\,M)}\,\left[
\frac{1}{2\,M^2}\sqrt{s\,(s - 4\,M^2)} \right.
$$
$$
\left. + \ln\, \frac{\sqrt{s} - \sqrt{s - 4\,M^2}}{\sqrt{s} +
\sqrt{s - 4\,M^2}}\right]\;,
$$
where $\kappa_u,\kappa_{\bar d}$ are anomalous magnetic moments of $u$ - and
$\bar d$-quarks.

It is of interest to separate the contribution to (\ref{mu}) of purely
relativistic kinematic effect of the Wigner rotation of quark spins
\cite{KrT02,KrT03} that appears under relativistic invariant spin
summation. If we put $\omega_1 = \omega_2$ = 0 in the free two-particle
magnetic form factor (see Appendix) we obtain the magnetic moment without
taking into account the Wigner spin rotation in the form:
$$
\tilde\mu_\rho = \frac{M_\rho}{2M}\,(1 + \kappa_u + \kappa_{\bar d})\,
$$
\begin{equation}
\times\int_{2M}^\infty\,d\sqrt{s}\, \frac{\varphi^2(s)} {\sqrt{s -
4\,M^2}}\,\left\{1 - \frac{1}{2}\,L(s)\right\}\;, \label{muwsr}
\end{equation}
The spin-rotation contribution is negative and its value is
$\approx 15\%$ of that of the $\rho$-meson magnetic moment.

To calculate the numerical value of the $\rho$-meson magnetic moment
(\ref{mu}) we use for the wave functions in the sense of RQM
(\ref{wfRQM}) the following model wave function that depends on only one
parameter and is largely used in relativistic composite quark model
(on a
level with harmonic oscillator wave function)
\cite{BrS95} (see, also \cite{CoP05}),
the power-law wave functions:
\begin{equation}
\psi(k) =N_{PL}\,{(k^2/b_\rho^2 + 1)^{-n}}\;,\quad n=2,3\;.
\label{PLwf}
\end{equation}

The parameters that enter our calculation in whole are from two groups:

1) the parameters that describe the constituent quarks {\it per
se} (the quark mass
$M$, the anomalous magnetic moments of quarks $\kappa_q$,
that enter our formulae through the sum
$s_q =
\kappa_u + \kappa_{\bar d}$, and the quark mean square radius (MSR) $\langle
r^2_q\rangle$);

2) the parameter  $b_\rho$ that enters the quark wave function
(\ref{PLwf}) and is determined by the quark interaction potential.

As it was shown in \cite{KrP16},
to calculate electroweak properties of
the $\rho$  meson
one can  use the same values of quark parameters from the first group as
that were used for the pion \cite{KrT01}.
In the paper \cite{KrT01} on pion we have shown that in our approach
all the parameters of the first group are the functions of the quark mass
$M$ and are defined by its value. In particular, for the quark MSR we can
use the relation (see, also \cite{CaG96}):
\begin{equation}
\langle r^2_q\rangle \simeq 0.3 /M^2\;.
\label{r2q}
\end{equation}

In our $\pi$\&$\rho$ model \cite{KrP16} the mass of the constituent
$u$- è $\bar d$ quarks was choosen to be $M =$ 0.22 GeV.
The sum of the anomalous magnetic moments of quarks was taken as
$s_q=$ 0.0268 in quark magnetons.
It is just these values of the parameters that in the frame of our version
of IF RQM provide the successful and predictive description of pion
properties
\cite{KrT01, KrT09prc},
as it was mentioned
in the Introduction.

For the parameter
$b_\rho$
of the wave function we use the value
obtained in our paper
\cite{KrP16} from the fitting of the lepton decay constant of the $\rho$
meson. Using the wave function (\ref{PLwf}) with $n=$ 3 we have obtained
$b_\rho = (0.385 \pm 0.019)$~GeV. The uncertainty is due to the
uncertainty of the experimental value of the decay constant
 $f_{\rho}^{\rm exp} =
(152 \pm 8)$~MeV \cite{MeS15,Oli14}.
The calculation of  \cite{KrP16} schematically can be presented as a
chain :
\begin{equation}
f_{\rho}^{\rm exp} \to b_\rho \to  \langle r^2_\rho\rangle
\label{chain}
\end{equation}
and the obtained value $\langle r^2_\rho\rangle = (0.56 \pm 0.04)$ fm$^{2}$
one can consider as a theoretical extraction of the experimental value
(see, e.g., \cite{BaK17}).

The electromagnetic form factors of constituent quarks are taken
in the form \cite{KrT09,Kru97,KrT98,TrT13},
$$
G^{q}_{E}(Q^2) = e_q\,f_q(Q^2)\;,
$$
\begin{equation}
G^{q}_{M}(Q^2) = (e_q + \kappa_q)\,f_q(Q^2)\;,
\label{q ff}
\end{equation}
where $e_q$ is the quark charge and $\kappa_q$ is the quark
anomalous magnetic moment. The quark form factor has the form \cite{Kru97}
\begin{equation}
f_q(Q^2) = \frac{1}{1 + \ln(1+ \langle r^2_q\rangle Q^2/6)}\; ,
\label{f_qour}
\end{equation}
$\langle r^2_q\rangle$ is the MSR of the constituent quark. Values
of all parameters used in these expressions are taken from the
$\pi$-meson calculation. As we have mentioned these values
give the pion form factor asymptotics at large momentum transfer that
coincides with that of QCD
(see e.g. Ref.~\cite{TrT13}).

In Figures 1 - 3 the results for the
electromagnetic $\rho$-meson form factors obtained with the parameters
described above are presented. For comparison the results from some other
papers are given. Our quadrupole form factor differs strongly from that of
other given results. In this context it is necessary to note that we
(as opposed to, e.g., \cite{ChJ04}) take into account the quark
structure in terms of quark form factors (\ref{q ff}), (\ref{f_qour}),
that accelerating the decay of form factors. Note that the calculation
using the formulae analogous to (\ref{GqGRIP}) described well the
electron-deuteron polarization scattering \cite{KrT07} and so the
experimental deuteron quadrupole form factor.

\begin{figure}[h!]
\epsfxsize=0.9\textwidth
\centerline{\psfig{figure=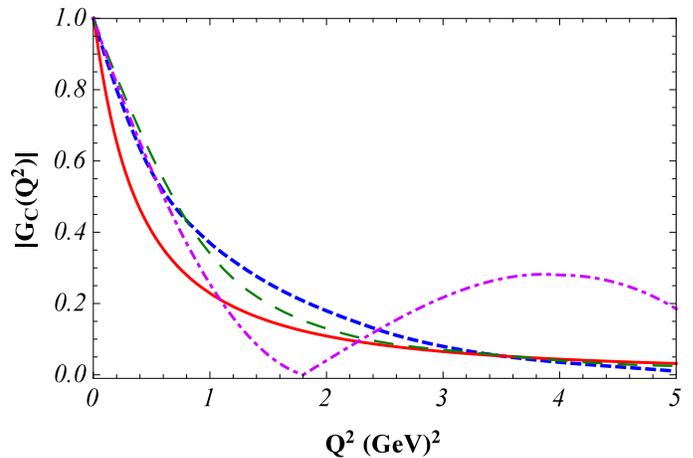,width=9cm}}
\vspace{0.3cm}
\caption{Modulus of the charge $\rho$ meson form factor. Solid line (red) -- the results of our calculations with wave function (\ref{PLwf}) at $n$=3 and parameters described in the text;
dashed line (green) -- the results of calculation in point-form dynamics \cite{BiS14};  short-dashed line (blue) --  the result of calculations in light-front dynamics \cite{ChJ04}; dot-dashed line (violet) -- the results of calculation in the Dayson-Schwinger equation approach \cite{HaP99}.}
\label{fig:1}
\end{figure}

\begin{figure}[h!]
\epsfxsize=0.9\textwidth
\centerline{\psfig{figure=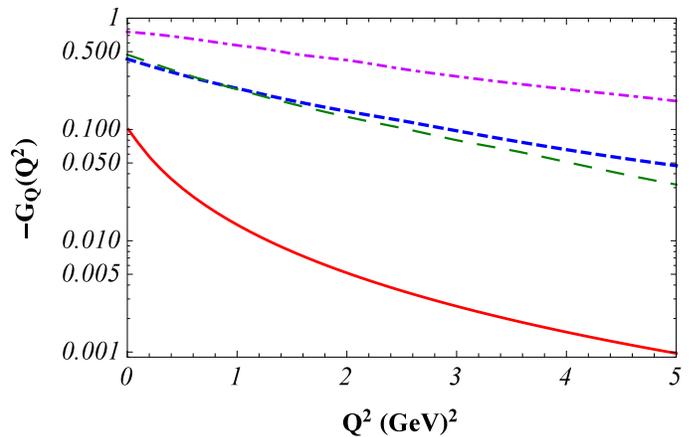,width=9cm}}
\vspace{0.3cm}
\caption{The quadrupole $\rho$ meson form factor in the logarithmic scale, legend as in Fig.\ref{fig:1}.}
\label{fig:2}
\end{figure}

\begin{figure}[h!]
\epsfxsize=0.9\textwidth
\centerline{\psfig{figure=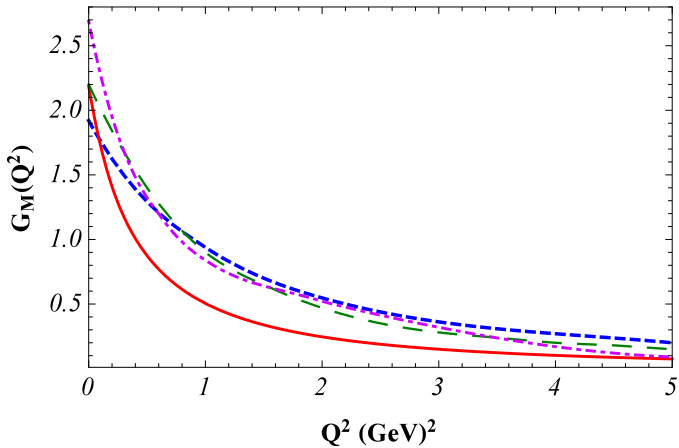,width=9cm}}
\vspace{0.3cm}
\caption{The magnetic  $\rho$ meson form factor, legend as in Fig.\ref{fig:1}.}
\label{fig:3}
\end{figure}

Finally, using the mentioned values of the parameters of the  $\pi$\&$\rho$
model, we obtain from (\ref{mu}) the following value of the magnetic moment
$$
\mu_{\rho} = 2.16 \pm 0.03\hspace{1mm}[e/2M_{\rho}]
$$
which is in
accordance with the conventional experimental data.

During the last decade, a number of papers appeared that considered the
electroweak properties of $\rho $ meson according to the growing interest
to the problem. A lot of theoretical information is obtained while the
experimental base remains scare. We list below only the papers where the
value of the $\rho $-meson magnetic moment is given whereas a lot of
interesting papers on  $\rho $ meson remain
out of scope, for example those based on the holographic approach
\cite{GrR07}.

First, we refer to different formulation of RQM in the frame of
relativistic composite model (see, e.g.,
\cite{CaG95,MeF97,BaC02,Jau03,ChJ04,HeJ04,BiS14,SuD17,MeF18}. These approaches
are the most close in spirit to our approach.
 There are approaches
based on the Dyson-Schwinger equation
\cite{HaP99,IvK99,BhM08,RoB11,PiS13}, on
the Nambu-Jona-Lasinio model
\cite{CaB15,LuC15}. Some authors use
QCD sum rules \cite{Sam03,AlO09},
Feynmann diagrams in the light front formalism
\cite{MeS02},  the bag model \cite{Sim16}, relativistic composite model
\cite{BaE85}, the constructed relativistic Hamiltonian
\cite{BaS13}, a low-energy effective field theory \cite{DjE14},
or lattice QCD calculations \cite{AnW97,HeK07,LeM08,OwK15,LuS16}.
In the paper \cite{GuS15} the $\rho$ meson magnetic moment is obtained
from \textit{BABAR} cross section data for the reaction $e^{+}e^{-} \to
\pi^{+}\pi^{-}2\pi^{0}$. They found the values using preliminary data from
the \textit{BABAR} Collaboration for the $\mu_{\rho} ^{exper}$ =$2.1 \pm
0.5$ [$e/2M_{\rho}$]. \\
In a recent paper, \textit{BABAR} Collaboration presents \cite{BaB17} the
new experimental results concerning the mentioned reaction. One can hope
that the new data processing would make it possible to increase the
precision of the value of $\mu_{\rho}$ extracted from the
experimental data.

 The values of the $\rho $-meson magnetic moment in different
 approaches are presented in Table I.

\begin{table}[h!]
\caption{The comparison of the results for the magnetic moment
$\mu_{\rho}$ (in natural magnetons $e/2M_{\rho}$) in different approaches
.}
\label{tab:1}
\begin{tabular}{|c|c|}
\hline
Model  & $\mu_{\rho}$\\
\hline
This work, mIF RHD  & $2.16 \pm 0.03$ \\
\hline
Cardarelly, LF RHD \cite{CaG95} & 2.26\\
\hline
Melo, LF RHD \cite{MeF97}&2.14 \\
\hline
Bakker, LF RHD \cite{BaC02} & 2.1\\
\hline
Jaus, LF RHD \cite{Jau03}& 1.83\\
\hline
 Choi, LF RHD \cite{ChJ04} & 1.92\\
\hline
He, LF, IF RHD \cite{HeJ04} & 1.5\\
\hline
He, PF RHD \cite{HeJ04} & 0.9\\
\hline
Biernat, PF RHD \cite{BiS14} & 2.20\\
\hline
Sun, LCCCM \cite{SuD17} & 2.06\\
\hline
Hawes, DSE \cite{HaP99} & 2.69\\
\hline
Ivanov, DSE \cite{IvK99} & 2.44\\
\hline
Bhagwat, DSE \cite{BhM08} & 2.01\\
\hline
Roberts, DSE \cite{RoB11} & 2.11\\
\hline
Pitschmann, DSE \cite{PiS13} & 2.11\\
\hline
Carrillo-Serrano, NJL \cite{CaB15} & 2.59\\
\hline
 Luan, NJL \cite{LuC15} & 2.1\\
\hline
Samsonov, QCD sum rules \cite{Sam03} & $2.0 \pm 0.3$\\
\hline
Aliev, QCD sum rules \cite{AlO09} & $2.4 \pm 0.4$ \\
\hline
Melikhov, LF triangle \cite{MeS02} &2.35\\
\hline
\v{S}imonis, bag model \cite{Sim16} & 2.06 \\
\hline
Bagdasaryan, relativistic CQM \cite{BaE85} & 2.3\\
\hline
 Badalian, RH \cite{BaS13} &1.96\\
\hline
Djukanovic, EFT \cite{DjE14} & 2.24\\
\hline
Andersen, Latt. \cite{AnW97} &$ 2.25 \pm 0.34$\\
\hline
Hedditch, Latt. \cite{HeK07} & 2.02 \\
\hline
Lee, Latt. \cite{LeM08} & $ 2.39 \pm 0.01$\\
\hline
Owen, Latt. \cite{OwK15} & $2.21 \pm 0.08$\\
\hline
Lushevskaya, Latt. \cite{LuS16} & $2.11 \pm 0.10$\\
\hline
Gudi$\tilde n$o, Exper. \cite{GuS15} & $2.1 \pm 0.5$\\
\hline
\end{tabular}
\end{table}

One can see from Table I that the majority of results is grouped
close to the value $\mu_{\rho} = 2 e/2M_{\rho}$ , that is near
the gyromagnetic ratio $g = 2$. The authors of the paper \cite{Ter16}
noted, that universal closeness  to $g = 2$ may be also understood in
comparison to the Equivalence Principle for vector mesons.

Let us discuss briefly our results.
We emphasize once again that our calculation of the magnetic moment in
the frane of the $\pi$\&$\rho $ - model is parameter-free and is in
a very good accordance with experiment.
The given value of the uncertainty of our result is totally due
to the uncertainty of the experimental data for
$f_{\rho}$. So, the refinement of this data will bring about the decrease
of the uncertainty of our result for
$\mu_{\rho}$ .

It is worth to notice that the $\pi$\&$\rho$ model gives the value of the
charge $\rho $- meson radius \cite{KrP16} in accordance with the Wu--Yang
hypothesis \cite{WuY65} (see also Refs.~\cite{ChY68, PoH87, PoH90, Gou74}).

\section{Conclusions}
\label{sec:concl}

To summarize, we calculate the
magnetic dipole moment of the $\rho$ meson, $\mu_{\rho}$,
in relativistic constituent-quark model.
We derive explicit analytical expression for
$\mu_{\rho}$ exploiting
our relativistic approach to composite systems,  a version of the
instant-form Relativistic  Quantum Mechanics.
We have used this approach in the paper \cite{KrP16}
 to construct
a unified $\pi $\&$\rho $ model
describing electroweak properties of
light mesons.
In the present paper we calculate $\mu_{\rho}$ using our
unified $\pi $\&$\rho $ model
without addition of any fitting parameters.
This parameter-free calculation  gives the value
contribution of the Wigner spin rotation to the magnetic moment
$\mu_{\rho}$ is ($\approx$ 15\%). We consider the small uncertainty
($\approx$ 1.4\%). of our value of magnetic moment as one of undoubted
advantages of the  method. A comparison is made with recent lattice QCD
results and other calculations using variety of methods.

So, we can state that in the frame of our $\pi$\&$\rho$ model, that is in
our version of IF RQM approach and at the same common values of
quark parameters, the concordant description of electroweak properties of
the $\pi $ and $\rho $ mesons is obtained. The constructed
relativistic approach demonstrated the predictability in describing the
pion electromagnetic form factors (see our paper \cite{KrT09prc}) and
gives, in accordance with experimental data, the electroweak
characteristics of the pion as well as of the $\rho$ meson.

\begin{acknowledgments}
One of the authors (V.T.) is grateful to S.V.Troitsky for helpful
discussions and valuable comments.
\end{acknowledgments}

\section*{Appendix}
\label{sec:Append}

The charge $g_{0C}$, quadrupole $g_{0Q}$ and magnetic $g_{0M}$ form factors for free two--particle system are:
$$
g_{0C}(s, Q^2, s') = \frac{1}{3}\,R(s, Q^2, s')\,Q^2
$$
$$
\times\left\{(s + s'+ Q^2)(G^u_E(Q^2)+G^{\bar d} _E(Q^2))\right.
$$
$$
\times\left[2\,\cos(\omega_1-\omega_2) + \cos(\omega_1+\omega_2)\right]
$$
$$
- \frac{1}{M}\xi(s,Q^2,s')(G^u_M(Q^2)+G^{\bar d}_M(Q^2))
$$
$$
\times\left.\left[2\,\sin(\omega_1-\omega_2) - \sin(\omega_1+\omega_2)\right]\right\};
\eqno{(A1)}
$$
$$
g_{0Q}(s, Q^2, s') = \frac{1}{2}\,R(s, Q^2, s')\,Q^2
$$
$$
\times\left\{(s + s'+ Q^2)(G^u_E(Q^2)+G^{\bar d}_E(Q^2))\right.
$$
$$
\times\left[\cos(\omega_1-\omega_2) - \cos(\omega_1+\omega_2)\right]
$$
$$
- \frac{1}{M}\xi(s,Q^2,s')(G^u_M(Q^2)+G^{\bar d}_M(Q^2))
$$
$$
\times\left.\left[\sin(\omega_1-\omega_2) + \sin(\omega_1+\omega_2)\right]\right\};
\eqno{(A2)}
$$
$$
g_{0M}(s, Q^2, s') = -\,{2}\,R(s, Q^2, s')\,
$$
$$
\times\left\{\xi(s,Q^2,s')\left[G^u_E(Q^2)+G^{\bar d}
_E(Q^2)\right]\, \sin(\omega_1-\omega_2) \right.
$$
$$
+ \frac{1}{4\,M}\left[G^u_M(Q^2)+G^{\bar d}_M(Q^2)\right]\,
\left[(s + s' +Q^2)\,Q^2\,\right.
$$
$$
\times\left(\frac{3}{2}\,\cos(\omega_1-\omega_2) +
\frac{1}{2}\cos(\omega_1+\omega_2)\right)
$$
$$
- \frac{1}{4}\xi(s,Q^2,s')
$$
$$
\times\left[\frac{(\sqrt{s'}+2\,M)(s-s'+Q^2)+(s'-s+Q^2)\sqrt{s'}}
{\sqrt{s'}(\sqrt{s'}+2\,M)}  \right.
$$
$$
\left. + \frac{(\sqrt{s}+2\,M)(s'-s+Q^2)+(s-s'+Q^2)\sqrt{s}}
{\sqrt{s}(\sqrt{s}+2\,M)}\right]
$$
$$
\times\left[\sin(\omega_1-\omega_2) -
\sin(\omega_1+\omega_2)\right]
$$
$$
- \frac{1}{2}\xi^2(s,Q^2,s')\left[
\frac{1}{\sqrt{s'}(\sqrt{s'}+2\,M)} +
\frac{1}{\sqrt{s}(\sqrt{s}+2\,M)}\right]
$$
$$
\left.\left. \times \left[\cos(\omega_1-\omega_2) +
\cos(\omega_1+\omega_2)\right]\right]\right\} \;. \eqno{(A3)}
$$
Here
$$
R(s, Q^2, s') = \frac{(s + s'+ Q^2)}{2\sqrt{(s-4M^2) (s'-4M^2)}}\,
$$
$$
\times\frac{\vartheta(s,Q^2,s')}{{[\lambda(s,-Q^2,s')]}^{3/2}}
\frac{1}{\sqrt{1+Q^2/4M^2}}\;,
$$
$$
\xi(s,Q^2,s')=\sqrt{ss'Q^2-M^2\lambda(s,-Q^2,s')}\;,
$$
$\omega_1$ and $\omega_2$ are the Wigner rotation parameters:
$$
\omega_1 =
\arctan\frac{\xi(s,Q^2,s')}{M\left[(\sqrt{s}+\sqrt{s'})^2 +
Q^2\right] + \sqrt{ss'}(\sqrt{s} +\sqrt{s'})}\;,
$$
$$
\omega_2 = \arctan\frac{ \alpha (s,s') \xi(s,Q^2,s')} {M(s + s' +
Q^2) \alpha (s,s') + \sqrt{ss'}(4M^2 + Q^2)}\;,
$$
here $\alpha (s,s') = 2M + \sqrt{s} + \sqrt{s'} $,
$\vartheta(s,Q^2,s')= \theta(s'-s_1)-\theta(s'-s_2)$, $\theta$ is
the step--function.
$$
s_{1,2}=2M^2+\frac{1}{2M^2} (2M^2+Q^2)(s-2M^2)
$$
$$
\mp \frac{1}{2M^2} \sqrt{Q^2(Q^2+4M^2)s(s-4M^2)}\;.
$$
$M$ -- the mass of $u$-- and $\bar d$ quarks. The functions
$s_{1,2}(s,Q^2)$ give the kinematically available region in the
plane $(s,s')$.  $G^{u,\bar d}_{E,M}(Q^2)$-- Sachs form factors of
$u$-- and $\bar d$ quarks.

\end{document}